\documentclass[12pt,a4paper]{article}
\usepackage{url}      %
\newcommand{\doi}[1]{\url{http://dx.doi.org/#1}}
\newcommand{\GCD}{{\rm GCD}}
\newcommand{\calC}{{\cal C}}
\newcommand{\xorc}{\;{\char'136}\;}			%
\newcommand{\xorequals}{\;{\char'136}{\char'075}\;}	%
\newcommand{\leftshift}{\;{\char'074}{\char'074}\;}	%
\newcommand{\rightshift}{\;{\char'076}{\char'076}\;}	%
\usepackage{amsmath}  %

\title{Some long-period random number generators using shifts and xors}

\author{
Richard.~P.~Brent\thanks{%
\hbox{MSI, ANU, Canberra, ACT~0200, \textsc{Australia}.}
\hspace*{\fill}
\hbox{Presented at CTAC06.} 
\hbox{To appear in {\em ANZIAM Journal}.}
\hspace*{\fill}						
\hbox{Copyright \copyright~2006, 2007 the author.}
\hspace*{\fill} rpb224		
}}	

\date{2 July 2007}				%

\begin{document}

\maketitle

\begin{abstract} %
Marsaglia recently introduced a class of ``xorshift'' random number
generators (RNGs) with periods $2^n-1$ for $n = 32, 64$, etc.  
Here Marsaglia's xorshift generators are generalised to obtain fast 
and high-quality RNGs with extremely long periods. Whereas RNGs based on
primitive trinomials may be unsatisfactory, because a trinomial has very
small weight, these new generators can be chosen so that their minimal 
polynomials have a large number of non-zero terms and, hence, a large 
weight.
A computer search using Magma has found good RNGs for $n$ a power of
two up to 4096. These RNGs have been implemented in a free software package 
{\tt xorgens}.
\end{abstract}

\tableofcontents

\section{Introduction}
\label{sec:intro}

Marsaglia~\cite{Marsaglia03} proposed a class of uniform random number
generators called ``xorshift RNGs''. Their implementation requires only a
small number of left shifts, right shifts and ``exclusive or'' operations
per pseudo-random number.

Assume that the computer wordlength is $w$ bits (typically $w = 32$ or $64$).
Marsaglia's xorshift RNGs have period $2^n - 1$, where $n$ is a small 
multiple of $w$, say $n = rw$.

The present author~\cite{rpb218} showed that Marsaglia's xorshift RNGs 
are a special case of the well-known linear feedback shift register (LFSR)
class of RNGs. This was also observed by 
Panneton and L'Ecuyer~\cite{Panneton05}.
However, the xorshift RNGs have implementation advantages because 
$n$ (the number of state bits) is a multiple of the wordlength $w$.
In contrast, for RNGs based on primitive trinomials,
the corresponding parameter $n$ can not be a multiple of eight
(due to Swan's theorem~\cite{Menezes,Swan}) and
is usually an odd
prime. For example, $n = 19937$ in the case of the 
``Mersenne twister''~\cite{Matsumoto98}.
The ``tempering'' step of the Mersenne twister
can be omitted in the xorshift RNGs. Thus, the xorshift RNGs are simpler
and potentially faster.

Any RNG based on a finite state must eventually cycle, but it is desirable 
for RNGs to have a very long period $T$ (the cycle length).  Most generators
fail certain statistical tests if more than about $T^{1/2}$ random numbers
are used~\cite{Panneton05}, or perhaps even about $T^{1/3}$ numbers for the 
``birthday spacings'' test~\cite{Marsaglia85}.
For generators satisfying linear recurrences such as the LFSR 
generators with period $2^n - 1$, there is a linear relationship between
blocks of $n+1$ consecutive bits, so the generator may fail
statistical tests that detect this linear relationship.  
Also, on a parallel machine we may want to use disjoint segments of the
cycle on different processors, and if this is done by starting with 
different seeds on each processor, we want the probability that two segments
overlap to be negligible.  For all these reasons it is important for $n$ to
be large. The generators that we describe below have $n$ as large as 
$4096$ which is enough, but not so large that the generators are
slowed down by memory accesses. This is possible if the computer's 
memory cache is smaller than $n$ bits.

Marsaglia's original proposal~\cite{Marsaglia03} discusses mainly
the case $n \le 64$, but an extension to larger $n$ is suggested,
The present author has implemented a generalisation
{\tt xorgens}~\cite{xorgens}
with $n \le 4096$, in particular we can choose any power of two
$n = 2^k$ for $6 \le k \le 12$.  The problem in going to larger $n$ is 
that we need to know the complete prime factorisation of $2^n - 1$ in order
to be sure that the generator's period is maximal. These factorisations
are known for all multiples of $32$ up to $1632$
and for certain larger~$n$, see~\cite{Cunningham}. 
If we restrict $n$ to powers of two then it is
sufficient to know the factorisations of certain Fermat numbers
$F_k = 2^{2^k} + 1$, since for example
\begin{equation*}
2^{4096} - 1 = (2^{2048}+1)(2^{2048}-1) = F_{11}(2^{2048}-1) = \cdots = 
F_{11}F_{10}F_{9} \cdots F_{1}F_{0}\,.
\end{equation*}
The factorisations of the Fermat numbers $F_0, \ldots,  F_{11}$ are
known~\cite{rpb161}.

Panneton and L'Ecuyer~\cite{Panneton05} tested Marsaglia's xorshift RNGs and
found certain deficiencies, but they did not find any significant problems
with our {\tt xorgens} generators for $n \ge 128$. 

In Section~\ref{sec:theory} we introduce some notation, summarise the
relevant theory, and describe the class of RNGs implemented in our 
{\tt xorgens}
package. In Section~\ref{sec:generators} we discuss criteria for the
selection of ``optimal'' generators in the class, and give specific examples
of optimal generators for various $n \le 4096$ and $w = 32, 64$.  Finally,
in Section~\ref{sec:improvements} we discuss a known weakness of xorshift
RNGs and mention some possible improvements.

\section{Notation and theory}
\label{sec:theory}

Let $F_2 = {\rm GF}(2)$ be the finite field with two elements $\{0,1\}$.
We usually write addition in $F_2$ as $+$, but we use $\oplus$ 
if it is necessary to distinguish it from normal integer addition.
If $0$ is interpreted as ``false'' and $1$ as ``true'', then
the field operations are ``exclusive or'' ({\tt xor} or $\oplus$)
and ``and'' ($\land$).
A computer word of $w$ bits can be regarded as a vector $x$
of length $w$ over $F_2$. 
We shall identify a bit-vector $x$ with the corresponding integer (and
vice-versa) when necessary.  %

Our RNGs generate pseudo-random bit-vectors $x$, but these easily give
pseudo-random unsigned integers $x \in [0, 2^w)$,
pseudo-random signed integers $x - 2^{w-1} \in [-2^{w-1}, 2^{w-1})$,
or (by a linear transformation) pseudo-random real numbers in $(0, 1)$. 

Unfortunately there are two conventions for bit-vectors $x$: 
Marsaglia~\cite{Marsaglia03} uses {\em row} vectors $x \in F_2^{1 \times w}$,
but Panneton and L'Ecuyer~\cite{Panneton05} use 
{\em column} vectors $x \in F_2^{w \times 1}$. 
We shall follow Marsaglia and take $x$ as a row vector.
(To convert to column vector notation, transpose all equations
involving vectors and matrices.)

Fix parameters $r > s > 0$ (the choice of these will be discussed below),
and consider the linear recurrence
\begin{equation}
x^{(k)} = x^{(k-r)}A + x^{(k-s)}B\,, \label{eq:rec1}
\end{equation}
where $x^{(k)} \in F_2^{1 \times w}$.
Here $A$ and $B$ are fixed matrices in $F_2^{w \times w}$. 
Given $x^{(0)}$, $\ldots$, $x^{(r-1)}$, the 
recurrence~(\ref{eq:rec1}) uniquely defines the sequence 
$(x^{(k)})_{k \ge 0}$.

Let $L \in F_2^{w \times w}$ be the {\em left shift} matrix
\begin{equation*}
L = \left(\begin{array}{cccc} 
	0	& 0	 & \cdots	& 0 \\ 
	1	& 0	 & \cdots	& 0 \\
	\vdots	& \ddots & \ddots 	& \vdots \\
	0	& \cdots & 1		& 0
\end{array}\right)		
\end{equation*}
such that 
\begin{equation*}
(x_1, \ldots, x_w)L = (x_2, \ldots, x_w, 0)\,.
\end{equation*}
Similarly, let $R = L^T$ be the {\em right shift} matrix such that
\begin{equation*}
(x_1, \ldots, x_w)R = (0, x_1, \ldots, x_{w-1})\,.
\end{equation*}

Marsaglia's idea is to take $A$ and $B$ as products of a small number of
terms such as $(I + L^a)$ and $(I + R^b)$. Specifically, let us take
\begin{equation*}
A = (I + L^a)(I + R^b) 
\end{equation*}
and
\begin{equation*}
B = (I + L^c)(I + R^d) 
\end{equation*}
for small positive integer parameters $a, b, c, d$.
Marsaglia~\cite[\S3.1]{Marsaglia03} omits the factor $I + L^c$;
we include it for reasons of symmetry, to increase the number
of possible choices (see~\S\ref{sec:generators}), and to improve
properties related to Hamming weight (see~\S\ref{sec:improvements}).

With our choice of $A$ and $B$, the recurrence (\ref{eq:rec1})
becomes
\begin{equation}
x^{(k)} = x^{(k-r)}(I + L^a)(I + R^b) + x^{(k-s)}(I + L^c)(I + R^d)\,.
\label{eq:rec2}
\end{equation}

Note that, if $x$ is a bit-vector of length $w$, then $xL^a$ is just
$x$ shifted left $a$ places ($xL^a = 0$ if $a \ge w$),
and $x(I + L^a)$ is the xor of $x$ and $xL^a$.
The operation 
\begin{equation*}
x \leftarrow x(I + L^a)
\end{equation*}
can be written in C as
\begin{equation*}
{\tt x = x \xorc (x \leftshift a)}
\end{equation*}
or more succinctly as
\begin{equation*}
{\tt x \xorequals x \leftshift a}
\end{equation*}
(here $x$ is represented in a computer word ${\tt x}$
which C treats as an unsigned integer).
Similarly 
$x \leftarrow x(I + R^b)$
can be written in C as
${\tt x \xorequals x \rightshift b}$,
and 
\begin{equation*}
x \leftarrow x(I + L^a)(I + R^b)
\end{equation*}
can be written in C as
\begin{equation*}
{\tt x \xorequals x \leftshift a\; ; \; x \xorequals x \rightshift b\;.}
\end{equation*}

The recurrence~(\ref{eq:rec1}) is best 
implemented using a ``circular array'', that is an array where the
indices are computed mod~$r$ (see~\cite[\S3.2.2]{Knuth2}),
unless $r$ is very small. %

It is well-known that we can write the recurrence~(\ref{eq:rec1})
as
\begin{equation}
(x^{(k-r+1)} | x^{(k-r+2)} | \cdots | x^{(k)}) =
(x^{(k-r)} | x^{(k-r+1)} | \cdots | x^{(k-1)})\calC\,,
\label{eq:rec3}
\end{equation}
where the {\em companion matrix} $\calC \in F_2^{n \times n}$
can be regarded as an $r \times r$ matrix of $w \times w$ blocks
(recall that $n = rw$).
For example, if $r = 3$ and $s = 1$, then 
\begin{equation*}
(x^{(k-2)} | x^{(k-1)} | x^{(k)}) =
(x^{(k-3)} | x^{(k-2)} | x^{(k-1)})
\left(\begin{array}{ccc}
0 & 0 & A\\
I & 0 & 0\\
0 & I & B\\
\end{array}\right)\,.
\end{equation*}

The period of the recurrence~(\ref{eq:rec3}) is $2^n-1$ if the
characteristic polynomial 
\begin{equation*}
P(z) = \det\left(\calC - zI\right)
\end{equation*}
is primitive over $F_2$. 
$P(z)$ is primitive if it is irreducible %
and the powers $z, z^2, z^3, \ldots, z^{2^n-1}$ are distinct
mod~$P(z)$.  %
To verify this, without checking $2^n-1$ cases, 
it is sufficient to show that $P(z)$ is irreducible and
\begin{equation*}
z^{(2^n-1)/p} \ne 1 \bmod P(z)
\end{equation*} 
for each prime divisor
$p$ of $2^n-1$: see Lidl~\cite{Lidl} or Menezes~\cite[\S4.5]{Menezes}.

Suppose that 
\begin{equation*}
P(z) = \sum_{j=0}^n c_jz^j\,.
\end{equation*}
{From} the Cayley-Hamilton theorem, $P(\calC) = 0$, so
\begin{equation*}
\sum_{j=0}^n c_j\,\calC^j = 0\,.
\end{equation*}
It follows from~(\ref{eq:rec3}) that
\begin{equation*}
(x^{(j)} | x^{(j+1)} | \cdots | x^{(j+r-1)}) =
(x^{(0)} | x^{(1)} | \cdots | x^{(r-1)})\calC^j\,,
\end{equation*}
so 
\begin{equation*}
\sum_{j=0}^n c_j x^{(k+j)} = 0\,.
\end{equation*}
This shows that the pseudo-random sequence
$x^{(k)}$ satisfies a linear recurrence over $F_2$. 
For a good random number generator it 
is important that the {\em weight} $W(P(z))$ of the polynomial $P(z)$,
i.e.~the number of nonzero coefficients $c_j$,
is not too small~\cite{rpb218,Ecuyer04,Panneton05}. 

\pagebreak[3]		%

\section{Optimal generators}
\label{sec:generators}

Suppose the wordlength $w$ and a parameter $r \ge 2$ are given, so 
$n = rw$ is defined. We want to choose positive parameters $(s, a, b, c, d)$
such that $s < r$ and the RNG obtained from the recurrence~(\ref{eq:rec2})
has full period $2^n-1$. Of the many possible choices of $(s, a, b, c, d)$,
which is best?
We give a rationale for making the ``best'' choice (or at least a reasonably
good one, since often many choices are about equally good).

\begin{enumerate}
\item 
\label{item:abcd}
Each bit in $x(I+L^a)(I+R^b)$ should depend on at least two bits
in $x$, that is each column of the matrix $(I+L^a)(I+R^b)$ should have
weight (number of nonzeros) at least two.  A necessary condition for this
is that $a + b \le w$.  Similarly, we require that $c + d \le w$.
\item
Repeated applications of the transformation $x \leftarrow x(I+L^a)(I+R^b)$
should mix all the bits of the initial $x$ (that is, after a large number of
iterations each output bit should depend on each of the input bits).  A
necessary condition for this is that $\GCD(a,b) = 1$.  Similarly, we require
that $\GCD(c,d) = 1$.
\item
\label{item:wlog}
If $(s,a,b,c,d)$ is one set of parameters, then
$(s,b,a,d,c)$ is associated with the same characteristic polynomial.
We can assume that $a \ge b$,
as otherwise we could interchange $a \leftrightarrow b$, $c \leftrightarrow d$
to obtain an equivalent RNG.   
\item
\label{item:cd}
So that the left shift parameters ($a$ and $c$) are not both greater than
the right shift parameters ($b$ and $d$) we also assume that $c \le d$.
\item
\label{item:delta}
In order that the bits in $x(I+L^a)(I+R^b)$ depend on bits as far away as
possible (to both left and right) in $x$, we want to maximise $\min(a,b)$. 
Similarly, we want to maximise $\min(c,d)$.  Thus, we try to maximise
$\delta = \min(a,b,c,d)$.
\item
\label{item:primitive}
As already discussed, once $(a,b,c,d)$ are fixed, we want to choose $s < r$ 
so that the generator has full period $2^n-1$.
\item
\label{item:maxweight}
Finally, in case of a tie (two or more sets of parameters satisfying the
above conditions with the same value of $\delta$), we choose the set
whose characteristic polynomial has maximum weight $W$.
\end{enumerate}

There might still be a tie, that is two sets of parameters satisfying the
above conditions, with the same $\delta$ and $W$ values. However, because
the weights $W$ are quite large 
(see Tables~\ref{table:32-bit}--\ref{table:64-bit}), 
this is unlikely and has not been observed. 

Criteria~\ref{item:abcd} and~\ref{item:delta} 
lead to a simple search strategy. From
criterion~\ref{item:abcd} we see that $\delta \le w/2$,
but criterion~\ref{item:delta} is to maximise~$\delta$. 
Thus, we start from $\delta = \lfloor w/2 \rfloor$ and decrease
$\delta$ by $1$ until we find a quadruple of parameters $(a,b,c,d)$ 
satisfying criteria~\ref{item:abcd}--\ref{item:cd}. This involves
checking $O((w/2 - \delta)^4)$ possibilities
since $(a,b,c,d) \in [\delta, w - \delta]^4$.
We then search for $s$
satisfying criterion~\ref{item:primitive} 
(this is the most time-consuming step).
There are $r-1$ possibilities to check for each quadruple $(a,b,c,d)$. 
If no $s$ satisfying criterion~\ref{item:primitive}
is found, we decrement $\delta$ and repeat the process. 
Once one satisfactory quintuple $(s,a,b,c,d)$ has 
been found, we need only check other quintuples $(s',a',b',c',d')$ with
the same $\delta$, and choose the best according to 
criteria~\ref{item:primitive} and~\ref{item:maxweight}.
We need only consider $s$ such that $\GCD(r,s) = 1$, since this is
a necessary (but not sufficient) condition for the characteristic
polynomial to be irreducible.

There might not be a solution satisfying all the 
conditions~\ref{item:abcd}--\ref{item:maxweight}.
The number of candidates $(s,a,b,c,d)$ is or order $rw^4$, that is 
$nw^3$ since $n = rw$.
The probability that
a randomly chosen polynomial of degree $n$ over $F_2$ is primitive is 
between $1/(n\log n)$ and $1/n$, apart from constant factors
~\cite{Lidl,Menezes}.
Thus, if our characteristic polynomials behave like random polynomials of
the same degree, we expect at least of order $w^3/\log n$ solutions.
For $w \ge 32$ we have always
been able to find a solution with $w/2-\delta \le 9$.
If $w$ is small, there may be no solution,
for example there is no solution for $w = 8$, $r = 6$.

The parameters for ``optimal'' random number generators 
with $n$ a power of two
(up to $n = 4096$) are given in Tables~\ref{table:32-bit}--\ref{table:64-bit}.
Parameters when $n$ is not a power of two are available from the author's
web site~\cite{xorgens}. %
The computations were performed using Magma~\cite{Bosma97}.
 
We do not recommend the RNGs with $n \le 128$ since they may fail the
matrix-rank test in the Crush testing package~\cite{Simard05,Panneton05}.
However, no problems have been observed while testing the RNGs with 
$n \ge 256$. 

\begin{table}
\centering
\caption{32-bit generators.} 			%
\label{table:32-bit}
~\\[0pt]
\begin{tabular}{|c|c|c|c|c|c|c|c|c|}
\hline
$n$  & $r$ & $s$& $a$& $b$& $c$& $d$&$\delta$&$W$ \\ 
\hline
64   & 2   & 1  & 17 & 14 & 12 & 19 & 12 & 31 \\ 
128  & 4   & 3  & 15 & 14 & 12 & 17 & 12 & 55 \\
256  & 8   & 3  & 18 & 13 & 14 & 15 & 13 & 109 \\
512  & 16  & 1  & 17 & 15 & 13 & 14 & 13 & 185 \\
1024 & 32  & 15 & 19 & 11 & 13 & 16 & 11 & 225 \\
2048 & 64  & 59 & 19 & 12 & 14 & 15 & 12 & 213 \\
4096 & 128 & 95 & 17 & 12 & 13 & 15 & 12 & 251 \\
\hline
\end{tabular}
\end{table}

\begin{table}
\centering
\caption{64-bit generators.} 			%
\label{table:64-bit}
~\\[0pt]
\begin{tabular}{|c|c|c|c|c|c|c|c|c|}
\hline
$n$  & $r$ & $s$& $a$& $b$& $c$& $d$&$\delta$&$W$ \\ 
\hline
128  & 2   & 1  & 33 & 31 & 28 & 29  & 28 & 65  \\
256  & 4   & 3  & 37 & 27 & 29 & 33  & 27 & 127 \\
512  & 8   & 1  & 37 & 26 & 29 & 34  & 26 & 231 \\
1024 & 16  & 7  & 34 & 29 & 25 & 31  & 25 & 439 \\
2048 & 32  & 1  & 35 & 27 & 26 & 37  & 26 & 745 \\
4096 & 64  & 53 & 33 & 26 & 27 & 29  & 26 & 961 \\
\hline
\end{tabular}
\end{table}

\section{Problems and improvements}
\label{sec:improvements}

The {\tt xorgens} class of RNGs 
are easy to implement since only simple operations
(left and right shifts and xors) on full words are required. Unlike RNGs
based on primitive trinomials, their characteristic polynomials have high
weight (see column ``$W$'' of 
Tables~\ref{table:32-bit}--\ref{table:64-bit}). 
Provided $n \ge 256$, they appear to 
pass all common empirical tests for 
randomness~\cite{Simard05,Diehard,Panneton05}.

However, the {\tt xorgens} class, like Marsaglia's xorshift class,
does have an obvious theoretical weakness.
For $x \in F_2^{1 \times w}$, 
define $||x||$ to be the \emph{Hamming weight} of $x$, that is the 
number of nonzero components of $x$. Then $||x-y||$ is the usual
\emph{Hamming distance} between vectors $x$ and $y$.
For random vectors $x \in F_2^{1 \times w}$, $||x||$ has a 
binomial distribution with mean $w/2$ and variance $w/4$.

Because the matrices $(I+L^a)$ and $(I+R^b)$ are sparse, they map 
vectors with low Hamming weight into vectors with low Hamming weight,
in fact $||x(I+L^a)|| \le 2||x||$, $||x(I+R^b)|| \le 2||x||$,
and consequently 
\begin{equation*}
||x(I+L^a)(I+R^b)|| \le 4||x||\,.
\end{equation*}
It follows that a sequence $(x^{(k)})$ generated using 
the recurrence~(\ref{eq:rec2}) satisfies
\begin{equation*}
||x^{(k)}|| \le 4\left(||x^{(k-r)}|| + ||x^{(k-s)}||\right)\,.
\end{equation*}
Thus, the occurrence of a vector $x^{(k)}$ with low Hamming weight
is correlated with the occurrence of low Hamming weights further back
in the sequence (with lags $r$ and $s$). A statistical test could be 
devised to detect this behaviour in a sufficiently large sample.
It is a more serious problem for the 32-bit generators than for the
64-bit generators, since the probability that a $w$-bit vector $x$
has Hamming weight $||x|| \le w/8$ is $1.0 \times 10^{-5}$ for $w = 32$,
but only $2.8 \times 10^{-10}$ for $w = 64$.

One solution, recommended by Panneton and L'Ecuyer~\cite{Panneton05},
is to include more left and right shifts in the recurrence~(\ref{eq:rec2}).
This slows the RNG down, but
not by much, since most of the time is taken by loads, stores, and other
overheads. Another solution, which we prefer, is to combine the output
of the xorshift generator with the output of a generator in a different
class, for example a {\em Weyl generator} which has the simple form
\begin{equation*}
w^{(k)} = w^{(k-1)} + \omega \bmod 2^w\,.
\end{equation*}
Here ``$+$'' means integer addition (mod $2^w$) and 
$\omega$ is some odd constant (a good choice is an odd integer
close to $2^{w-1}(\sqrt{5}-1)$).
The generators in our {\tt xorgens} package return
\[w^{(k)}(I \oplus R^{\gamma}) + x^{(k)} \bmod 2^w\] 
instead of simply $x^{(k)}$. Here $\gamma \approx w/2$ is a constant.
This is better than returning $w^{(k)} + x^{(k)} \bmod 2^w$ (as was done in
an earlier version of {\tt xorgens}) because the least significant bit of
$w^{(k)}$ has period $2$, but all bits of $w^{(k)}(I \oplus R^{\gamma})$ 
have a longer period (about $2^{w/2}$), and this period is relatively prime to the period 
$2^n-1$ of $x^{(k)}$. Thus each bit in the output should have high
linear complexity~\cite{Menezes}.

Note that addition mod $2^w$ is not a linear operation on vectors over
$F_2$, so we are mixing operations in two algebraic structures. This is 
generally a good idea because it avoids regularities associated with
linearity. For example, suppose we use one of Marsaglia's xorshift generators
to initialise our state vector, and we do it three times with seeds
$s, s', s''$ satisfying $s = s' \oplus s''$; then by linearity over $F_2$
our three sequences $x, x', x''$ 
satisfy $x = x' \oplus x''$, which is clearly
undesirable.  This problem vanishes if the xorshift RNG used for
initialisation is modified by addition (mod $2^w$) of a Weyl generator,
as is done in the {\tt xorgens} package.

\section{Conclusions}

We have shown how Marsaglia's xorshift RNGs can be generalised to give
high-quality RNGs with extremely long periods (greater than
$10^{1232}$). 
The RNGs are fast and 
easy to implement because only word-aligned operations are
used and no ``tempering'' step~\cite{Matsumoto98} is required.
We discussed a potential problem related to correlation of
outputs with low Hamming weights, and showed that this can be overcome by
the simple expedient of combining the output of a generalised xorshift RNG
with the output of a Weyl generator. 
An implementation of the resulting RNG is available in a
free software package {\tt xorgens}~\cite{xorgens}.

\end{document}